# A robust watermarking algorithm for medical images

**Ahmed Nagm[1], Mohammed Safy[2]**
[1]1Electrical Engineering, Modern Academy, Egypt
[2]Electrical Engineering, Egyptian Academy for Engineering & Advanced Technology, Egypt

**ABSTRACT** (10 PT)

Integrated healthcare systems require the transmission of medical images between medical centres. The presence of watermarks in such images has become important for patient privacy protection. However, some important issues should be considered while watermarking an image. Among these issues, the watermark should be robust against attacks and does not affect the quality of the image. In this paper, a watermarking approach employing a robust dynamic secret code is proposed. This approach is to process every pixel of the digital image and not only the pixels of the regions of non-interest at the same time it preserves the image details. The performance of the proposed approach is evaluated using several performance measures such as the Mean Square Error (MSE), the Mean Absolute Error (MAE), the Peak Signal to Noise Ratio (PSNR), the Universal Image Quality Index (UIQI) and the Structural Similarity Index (SSIM). The proposed approach has been tested and shown robustness in detecting the intentional attacks that change image, specifically the most important diagnostic information.

***Corresponding Author:***

Mohammed Safy,
Electrical Engineering,
El-Nahda, Al Salam First, Cairo Governorate.
Email: msafy@eaeat.edu.eg

## 1. INTRODUCTION

Nowadays, transferring medical images between medical centers is necessary for exchanging knowledge and experiences. It also makes patient follow up by different centers while moving is possible. Moreover, it helps for continuing patient research studies even if the patient moves to a new different medical center [1]. As a result of the rapid development in the field of image processing and the high ability to change some of the information is only discovered by specialists. There was a need to encrypt the medical information sent to be preserved [2-4].

When sending the medical images between the medical centers, the main priority is to protect the patient information from unauthorized manipulation. Therefore, electronic medical centers give high priority to the transfer of images of the patient safely [5, 6].

Digital watermarking is considered as one of the best solutions of the above issue. By inserting special information, called a watermark, watermarking can enhance the security of medical images. The medical image is divided into two main parts. The first part is the Region of Interest (ROI) which contains the detailed information of the patient and any bit distortion of this part will lead to undesirable treatment for patient. The second part is the Region of Non-Interest (RONI) which can be used in the Watermarking processing. As a result of this division the watermark processing has two main steps one of them extracts the ROI from the medical image and the other applies the watermarking on RONI. Segmentation is the first step used to avoid the distortion in the ROI [7and 8]. The watermarking techniques are classified into two different domains.
The spatial-domain watermark insertion, it is easy and to simple implement, but it is not efficient against various noise and attacks. The other one is the transform domain; it is efficient and more robust against attacks.

In Jing Liu.2019 [9], based on the DTCWT-DCT and a henon Map a new, robust, multi-watermarking algorithm for medical images is proposed, which combines the transform domains perception of the low frequency coefficient invariance, the zero-watermarking concept and the encryption technology.it reduces the complexity of the algorithm and increases the watermark's capacity to embed medical images by using a robust multi-watermark.

In Emad Nabil et al. 2016 [10], enhanced the variation of the Flower Pollination Algorithm (FPA). It combines the Clonal Selection Algorithm (CSA) and the standard FPA and tried the new calculation by applying it to 23 improvement benchmark issues.

In T. Avudaiappan [11], the dual encryption procedure such as signcryption algorithms and blowfish are used to implement the security model for medical images. In the encryption process, using OFP based signcryption technique the Public key and the private key are optimized.

In Aleš Ro_ek [12], proposed a method that combines three fundamental approaches RONI, Zero Reversible watermarking. By water marking It succeed to secure the medical image without any changes in the patient medical information.

In Abhilasha Sharma.2015 [13], based on DWT and DCT to embed multiple watermarks into the cover image a water marking algorithm is proposed. To secure the medical information, the method used multiple watermarking in the form of image and text. The medical image is taken as cover image, the less robust logo image is embedded to the ROI region and the more robust and confidential EPR data is embedded into NROI region and of the cover image. In addition, the RSA method is used to encrypt the EPR data before embedding it into the cover.

To deal with this issue, an algorithm is proposed to product a secret code is called Originality Identifier Code (OIC) has a dynamic property and embed in the digital images with manner that is not noticeable and at the same time distributed on every pixel of the digital image. The proposed algorithm is divided into two phases: Creation phase which the digital object is prepared to be original protected and then study the Image Quality based on assessment metrics.

## 2. PROPOSED METHODOLOGY

In this method, an image is decomposed into three components using Color Filter array (CFA). One of these components is elected for further processing. This elected component is encrypted using a dynamic keys approach. These dynamic keys combine the patient ID, the name and the arrival date. After the encryption is carried out, the new component is called a modified comment. Next a substitution processing is performed in the frequency domain between one of the other CFA outputs and the modified component to get the final encrypted component. In the last stage, the demos icing algorithm is performed on the final encrypted component and the other two CFA outputs. This has been done in order to get an image that includes our proposed Originality Identifier Code (OIC). For illustration, the whole structure of the proposed algorithm is shown in Figure 1.

For medical images there are three main categories of watermarking algorithms, based on RONI [14-19], based on classical conventional digital watermarking [20-24] and based on reversible watermarking [25-28]. For the algorithm based on the first and second categories, the watermarking strategies degrade the image quality themselves. For the third category, although the ability of these algorithms to retrieve the images without distortion but they are not efficient.

There are two main categories of attacks, the Conventional Attacks that include the Gaussian Noise Attacks, the JPEG Attacks [29-30 and 31] and the Median Filter Attacks. And the Geometrical Attacks that include the rotation, the scaling, the translation and the Cropping Attacks. The main target of the traditional algorithms is how to restore the original image. But the proposed algorithm checks whether the patient's information has been deliberately changed or not. Figure 2 demonstrates the proposed algorithm that detects the attacks

Starting with the input color image I has a $n * m$ dimension as a RGB image with JPEG format. Where, it represented as a:

$$I = (Irc) = \begin{bmatrix} I_{11} & \cdots & I_1m \\ \vdots & \ddots & \vdots \\ In_1 & \cdots & Inm \end{bmatrix} \quad (1)$$

The represented matrix of input RGB image at (1) is composed from three matrixes have the same number of rows and columns of the input images. The first matrix is the Red component, the second matrix is the Green component and the third matrix is the Blue component as shown in (2).

$$R = (Rrc) = \begin{bmatrix} R_{11} & \cdots & R_1m \\ \vdots & \ddots & \vdots \\ Rn_1 & \cdots & Rnm \end{bmatrix}$$
$$G = (Grc) = \begin{bmatrix} G_{11} & \cdots & G_1m \\ \vdots & \ddots & \vdots \\ Gn_1 & \cdots & Gnm \end{bmatrix}$$
$$B = (Brc) = \begin{bmatrix} B_{11} & \cdots & B_1m \\ \vdots & \ddots & \vdots \\ Bn_1 & \cdots & Bnm \end{bmatrix} \quad (2)$$

Then for every pixel, there are athree numeric values in terms of Red, Green and Blue component as shown in (3).

$$(Irc) = \{(Rrc), (Grc), (Brc)\} \quad (3)$$

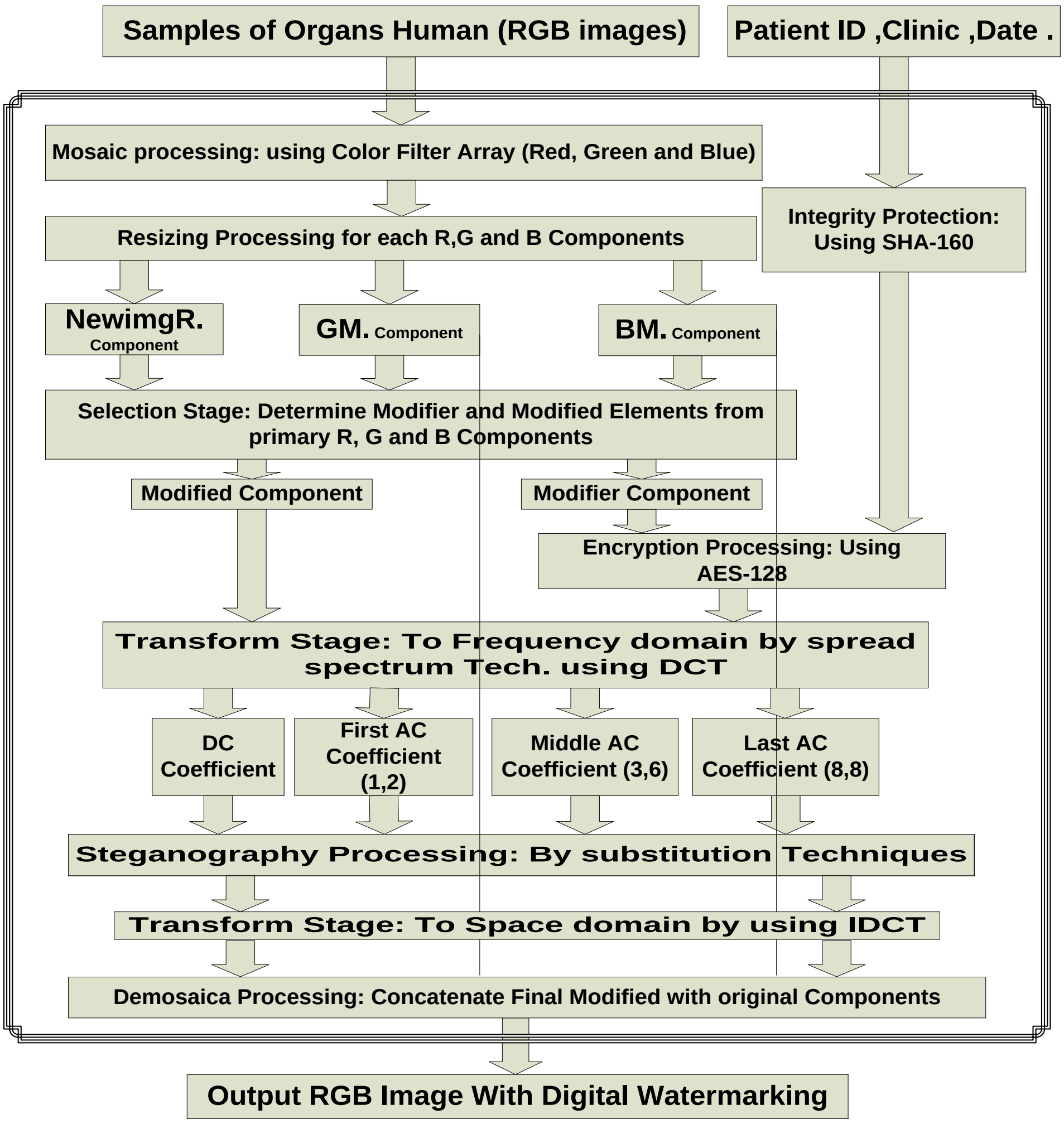

Figure 1. Proposed approaches of embedding digital watermarking into medical images.

The (4) shows the separation of the input image I to the three components.

$$R = I - (G + B)$$
$$G = I - (R + B)$$
$$B = I - (R + G) \quad (4)$$

Resize the separated Red, Green and Blue components to be suitable to the encryption algorithm.

$$\mathrm{RM(r)} = \mathrm{R}\left(\mathrm{r} + \left(8 - (\mathrm{r\ MOD\ 8})\right)\right), \text{if remender of numeric value for input rows} \neq 0$$
$$\text{else}, \mathrm{RM(r)} = \mathrm{R(r)} \quad (5)$$
$$\mathrm{RM(c)} = \mathrm{R}\left(\mathrm{c} + \left(8 - (\mathrm{c\ MOD\ 8})\right)\right), \text{if remender of numeric value for input colum} \neq 0$$
$$\text{else}, \mathrm{RM(c)} = \mathrm{R(c)} \quad (6)$$

The numerical intensity of the component depends on the used bits according to:

$$(\mathrm{Rrc}) = 2^{\mathrm{L}} - 1 \quad (7)$$

Where L: is the length of used bits, if L = 8 then the bits will be from 0 to 7. That mean the representation of pixels have intensities level from 0 (black) to 255 (white). And it is formatted as a depicted in (8).

$$(\mathrm{Rrc})_{10} = (\mathrm{Rrc})_2 = [\mathrm{b}_7, \mathrm{b}_6, \mathrm{b}_5, \mathrm{b}_4, \mathrm{b}_3, \mathrm{b}_2, \mathrm{b}_1, \mathrm{b}_0] \quad (8)$$

Where b: is the bit value (0 or 1).

The transformation function as a Discrete Cosine Transform is used, so (9) for each 8*8 block as following:

$$\mathrm{RM}^{\mathrm{t}}(\mathrm{u}, \mathrm{v}) = \left(\frac{2}{\mathrm{m}*\mathrm{n}}\right) \sum_{\mathrm{r}=0}^{\mathrm{m}-1} \sum_{\mathrm{c}=0}^{\mathrm{n}-1} \mathrm{R(r,c)} \left(\cos\left(\frac{(2\mathrm{r}+1)\pi \mathrm{u}}{2\mathrm{m}}\right) * \cos\left(\frac{(2\mathrm{c}+1)\pi \mathrm{v}}{2\mathrm{n}}\right)\right) \quad (9)$$

Where

| | |
|---|---|
| u | is raw number in the transformed data |
| v | is column number in the transformed data |
| r | is raw number in the original data |
| c | is column number in the original data |
| m | is number of rows in the original data |
| n | is number of columns in the original data |
| $\mathrm{RM}^{\mathrm{t}}$ | is value of pixel in the transformed data |
| RM | is value of pixel in the original data. |

$$\mathrm{RM}^{\mathrm{t}} = (\mathrm{Ruv}) = \begin{bmatrix} \mathrm{RM}^{11} & \cdots & \mathrm{RM}^{1}\mathrm{m} \\ \vdots & \ddots & \vdots \\ \mathrm{RMn}^{1} & \cdots & \mathrm{RMnm} \end{bmatrix} \quad (10)$$

To get the initial Key for the Symmetric Encryption Algorithm one of the crypto- graphic hash function on ID of capture device (Camera ID, as an example Serial number of capture device) as a SHA-1 is used to.

Performing the secure hash function on Capture ID by SHA-160, the output block size has 160 bit (Message Digest Length = 160). With initial input Key was “acef” and initial hash value were:

H [0] ='67452301',
H [1] ='EFCDAB89',
H [2] ='98BADCFE',
H [3] ='10325476',
H [4] ='C3D2E1F0',

Accordingly to (11), then the output will be:

$$\mathrm{K1} = \mathrm{SHA} - 1\ (\text{Camera ID}) = \mathrm{86E152C142DB1256FC1EF004ADEB7B935741D94D} \quad (11)$$

The first sixteen digits from the output is extracted to be the Initial key for encryption processing by AES-128 as shown in (12).

The Key K = '86E152C142DB1256' and CB are representing a ciphered blue component of original color image with forty-four internal key and 80 round according to (12).

$$\mathrm{CBrc} = \mathrm{AES} - 128([\mathrm{K}], (\mathrm{BMrc})) = \begin{bmatrix} \mathrm{CB}^{11} & \cdots & \mathrm{CB}^{1}\mathrm{m} \\ \vdots & \ddots & \vdots \\ \mathrm{CBn}^{1} & \cdots & \mathrm{CBnm} \end{bmatrix} \quad (12)$$

The steganography processing is started to get final modified component NRM in equation (13).

$$NRM = (NRMrc \Leftarrow CBrc) = \begin{bmatrix} NRM_{11} & \cdots & NRM_{1}m \\ \vdots & \ddots & \vdots \\ NRMn_{1} & \cdots & NRMnm \end{bmatrix} \Leftarrow \begin{bmatrix} CB_{11} & \cdots & CB_{1}m \\ \vdots & \ddots & \vdots \\ CBn_{1} & \cdots & CBnm \end{bmatrix} \quad (13)$$

The Discrete Cosine Transform substitution processing is performed on frequency domain between the element matrix of the ciphered blue and the red component and then the inverse transformation method is used to get the final modified component in the space domain.

The element number (1, 1) at every 8*8 block from the transformed red component is replaced by the same element of transformed ciphered blue component to get the final modified component at frequency domain then perform the inverse transform to get it at space domain.

At the end the modified red component is appeared according to (14).

$$NRM = (Ruv) = \begin{bmatrix} CB_{11} & \cdots & RM_{18} \\ \vdots & \ddots & \vdots \\ RM_{81} & \cdots & RM_{88} \end{bmatrix} \quad (14)$$

First AC coefficient, the element number (1, 2) at every 8*8 block from the transformed red component is replaced by the same element of transformed ciphered of the blue component to get the final modified component at frequency domain then perform the inverse transform to get it at the space domain, where the element $(1, 2)$ of $CB^t$ and other 64 coefficients for $RM^t$.

Middle AC coefficient, the element number (3, 6) at every 8*8 block from the transformed red component is replaced by the same element of the transformed ciphered of the blue component to get the final modified component at frequency domain then perform the inverse transform to get it at the space domain, where the element $(3, 6)$ of $CB^t$ and other 64 coefficients for $RM^t$.

Last AC coefficient, the element number (8, 8) at every 8*8 block from the transformed red component is replaced by the same element of the transformed ciphered of the blue component to get the final modified component at frequency domain then perform the inverse transform to get it at the space domain, where the element $(8, 8)$ of $CB^t$ and other 64 coefficients for $RM^t$.

At the end the modified red component will be according to the (15).

$$NRM = (RMuv) = \begin{bmatrix} RM_{11} & \cdots & RM_{18} \\ \vdots & \ddots & \vdots \\ RM_{81} & \cdots & CB_{88} \end{bmatrix} \quad (15)$$

The final modified component is concatenated with the two other original components “Modified Red, Green and Blue” accordingly to (16), where NI is protected originality image.

$$(NIrc) = \{(NRMrc), (GMrc), (BMrc)\} \quad (16)$$

There are two main classes to assist the image distortion one of them the mathematically class which includes MSE. MAE and PSNR and the other class is the human visual system.

The Mean Absolute Error and the Mean Square Error (MSE) are used to calculate the difference between the original image and the sent one.

$$MSE = \frac{1}{M*N} * \sum_{j=1}^{n} [x(i, j) - v(i, j)]^2 \quad (17)$$

$$MAE = \frac{1}{N} \sum_{i=1}^{N} |x_i - v_i| \quad (18)$$

The peak signal to noise ratio is used and it is used to measure the similarity between the input image and the image after watermarking.

$$PSNR = 10 \log_{10}\left(\frac{S^2}{MSE}\right) \quad (19)$$

Beside these metrics there are some important metrics like the Universal Image Quality Index (UIQI) which three factors loss of correlation, luminance distortion, and contrast distortion Also The most important metric is the Structural Similarity Index (SSIM) because it is not only measure the similarity between original image and the output image as a number but also it measures the similarity in the structure.

$$Q = \left(\frac{\sigma_{xv}}{\sigma_x \sigma_v}\right) * \left(\frac{2 * \bar{x} * \bar{v}}{\bar{x}^2 + \bar{v}^2}\right) * \left(\frac{2 * \sigma_x * \sigma_v}{\sigma_x^2 + \sigma_v^2}\right) \quad (20)$$

$$SSIM(X, V) = (2 * \bar{x} * \bar{v} + c1) * \frac{(2 * \sigma_{xv} + c2)}{(\bar{x}^2 + \bar{v}^2 + c1) * (\sigma_x^2 + \sigma_v^2 + c2)} \quad (21)$$

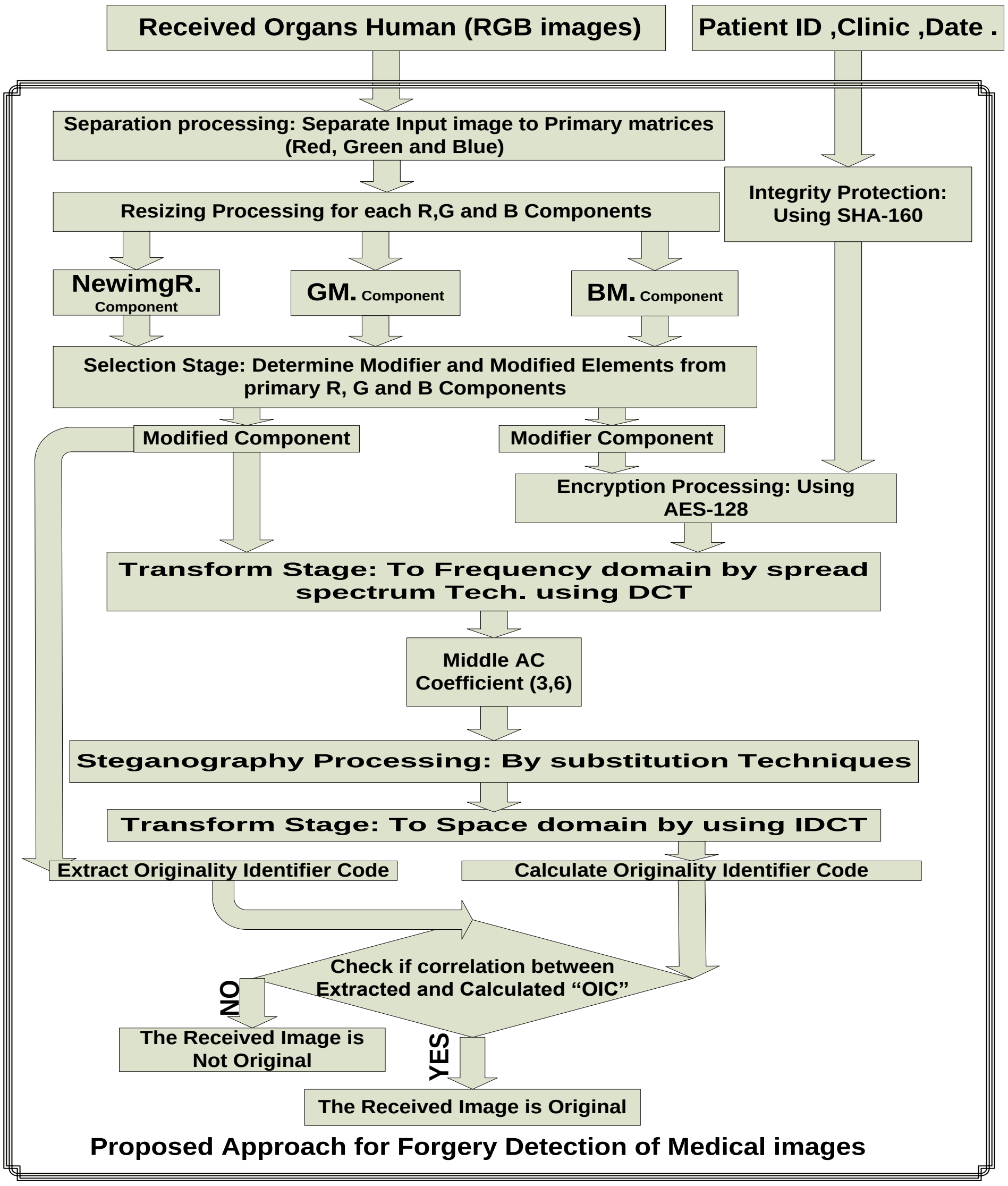

Figure 2. Proposed approch for forgery detection of medical images

## 3. RESULT AND ANALYSIS

In the experiment, MATLAB 2015 Version 8.5.0.197613, core i3 processor 2.3 Ghz and 4GB RAM is used as the test platform. In figure [2, 3, 4,5, 6 and 7] randomly selected medical images are used as an original image. PSNR, MAE, MSE, SSIM and UIQI are used to measure the performance of the proposed algorithm.

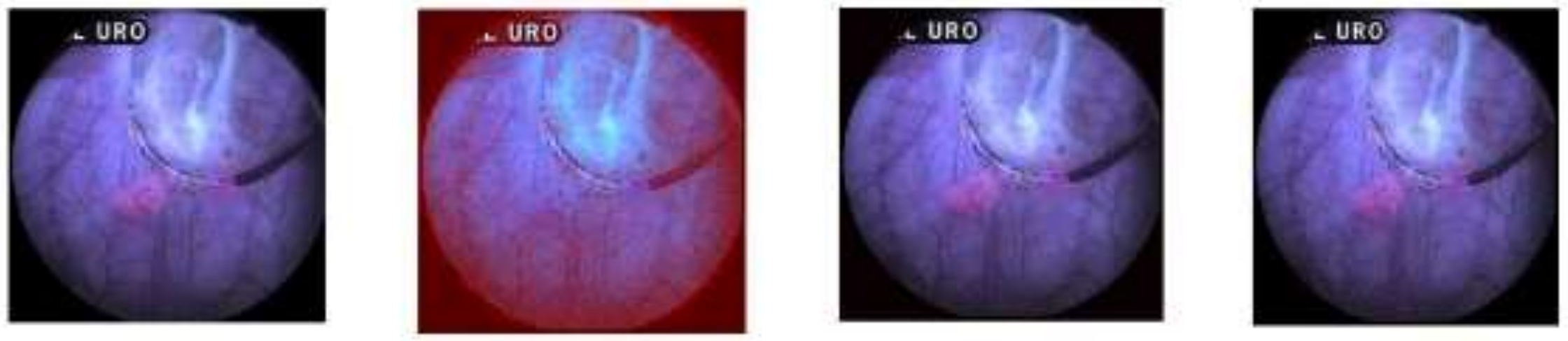

AA image without digital watermarks | AA image with digital watermarks by DC | AA image with digital watermarks by LAC | AA image with digital watermarks by FAC

Figure 2. Digital watermarking using different strategy for the AA image

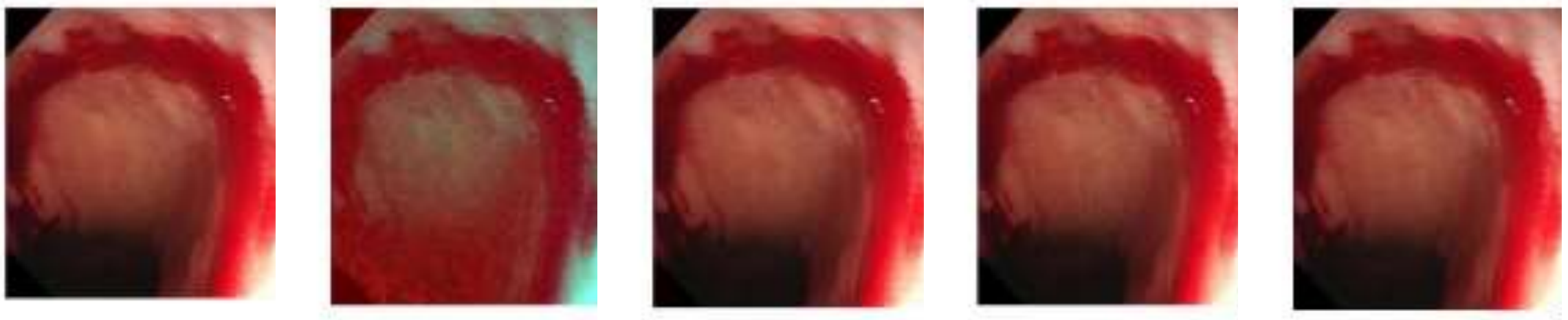

AB image without digital watermarks | AB image without digital watermarks by DC | AB image without digital watermarks by LAC | AB image without digital watermarks by FAC | AB image without digital watermarks by MAC

Figure 3. Digital watermarking using different strategy for the AB image

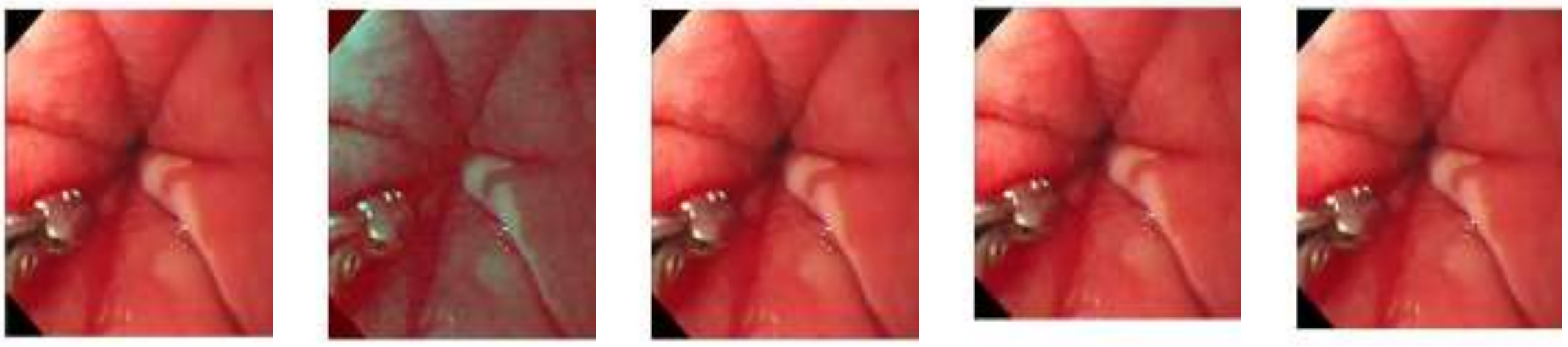

AC image without digital watermarks | AC image without digital watermarks by DC | AC image without digital watermarks by LAC | AC image without digital watermarks by FAC | AC image without digital watermarks by MAC

Figure 4. Digital watermarking using different strategy for the AC image

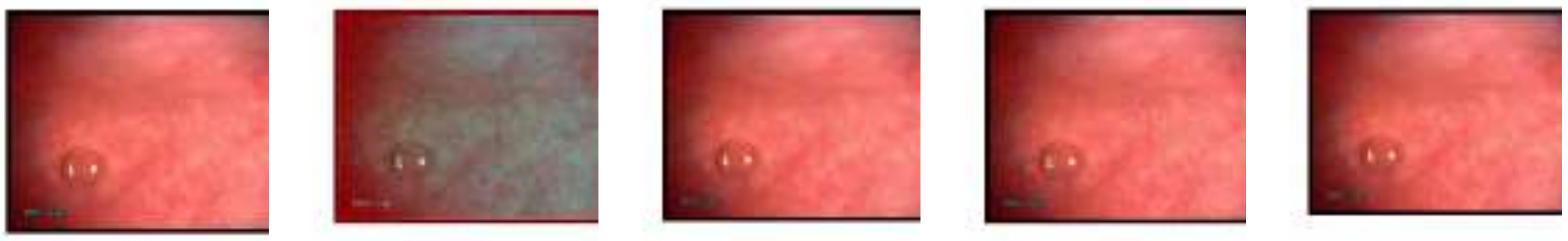

Flu image without digital watermarks | Flu image without digital watermarks by DC | Flu image without digital watermarks by LAC | Flu image without digital watermarks by FAC | Flu image without digital watermarks by MAC

Figure 5. Digital watermarking using different strategy for the Flu image

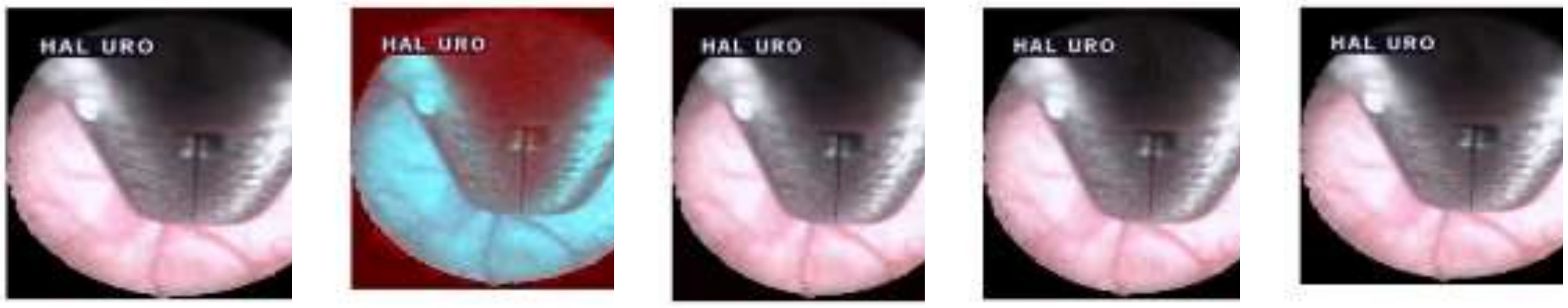

Thr image without digital watermarks | Thr image without digital watermarks by DC | Thr image without digital watermarks by LAC | Thr image without digital watermarks by FAC | Thr image without digital watermarks by MAC

Figure 6. Digital watermarking using different strategy for the Thr image

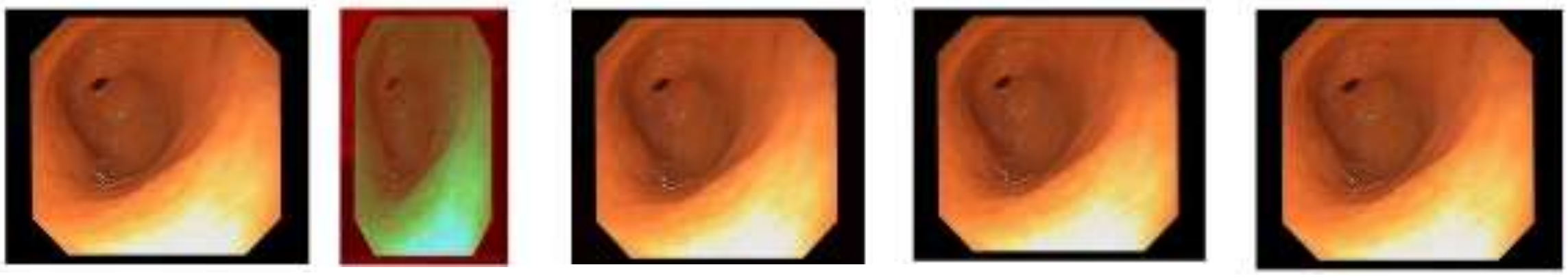

Gastro image without digital watermarks | Gastro image without digital watermarks by DC | Gastro image without digital watermarks by LAC | Gastro image without digital watermarks by FAC | Gastro image without digital watermarks by MAC

Figure 7. Digital watermarking using different strategy for the Gastro image

Table 1 represents the Mean Absolute of the proposed approach; the results show that the mean absolute error of the single MAC element (3, 6) has the lowest value compared with the other elements. Table 2 represents the Mean absolute error of the proposed approach; the results show that the mean square error of the single MAC element (3, 6) has the lowest value compared with the other elements. Table 3 represents the peak signal to noise ratio of the proposed approach; the results show that the peak signal to noise ratio of the single MAC element (3, 6) and single LAC element (8, 8) has the highest value compared with the other elements. Table 4 represents the Structural Similarity Index of the proposed approach; the results show that the Structural Similarity Index of the single MAC element (3, 6) and single FAC element (1,2) has the highest value compared with the other elements. Table 5 represents the Entropy the proposed approach; the results show that the Entropy of the single MAC element (3, 6), the single LAC element (8, 8) and FAC element (1,2) has the highest value compared with the DC element (1,2).

From analyzing these results, it is clear that the choosing of single MAC element (3, 6) gave the best performance over the rest of elements.

The vulnerability assessment approach is based on assuming an attack on one of the quadrants zones. In this approach, the technique should show full immunity against space attacks, whereas the image could be changed in any of the areas explained. Moreover, the assessment technique shall include as well the color component changes since the algorithm is based on manipulation based on color components. In this approach, and as explained in figure 8, the attack takes place either using a single-color basic component (Red, Green or Blue) or a natural color change (i.e. a mix of basic color components). Figures [9-13] demonstrate a different attack region and all of them are detected the proposed algorithm.

Table1. Mean absolute error of the proposed approach

| Images/Size/ JPEG | Mean Absolute Error of the Proposed Approach | | | |
|---|---|---|---|---|
| | Single-DC" Element (1,1) | Single-FAC" Element (1,2) | Single-MAC" Element (3,6) | Single-LAC" Element (8,8) |
| Thr/472*499, 21.9KB | 12.7863 | 1.1034 | 0.688 | 1.2792 |
| AA/459*442, 21.9KB | 16.2208 | 1.0384 | 0.8731 | 1.2973 |
| AB/512*512, 22KB | 6.624 | 1.0451 | 0.8251 | 0.9707 |
| AC/512*512, 340KB | 1.3006 | 1.1556 | 0.9568 | 0.9888 |
| Flu/348*288, 26KB | 3.7528 | 1.0867 | 0.8294 | 0.8301 |
| Gastro/640*480, 47.1KB | 9.8801 | 1.0979 | 0.6837 | 1.179 |

Table 2. Mean square error of the proposed approach

| Images/Size/ JPEG | Mean Square Error for the Proposed Approach | | | |
|---|---|---|---|---|
| | Single-DC" Element (1,1) | Single-FAC" Element (1,2) | Single-MAC" Element (3,6) | Single-LAC" Element (8,8) |
| Thr/472*499, 21.9KB | 2620.141746 | 52.14900149 | 18.58216191 | 35.56916981 |
| AA/459*442, 21.9KB | 1.29E+03 | 32.31012674 | 23.24093833 | 34.13997717 |
| AB/512*512, 22KB | 1.71E+03 | 32.49317678 | 23.22633489 | 29.98714193 |
| AC/512*512, 340KB | 1.60E+03 | 38.49172719 | 27.13516998 | 28.04392751 |
| Flu/348*288, 26KB | 3.09E+03 | 34.58821313 | 22.86935161 | 23.10755148 |
| Gastro/640*480, 47.1KB | 2620.141746 | 52.14900149 | 18.58216191 | 35.56916981 |

Table 3. The peak signal to noise ratio of the proposed approach

| Images/Size/ JPEG | Peak signal to noise ratio of the Proposed Approach | | | |
|---|---|---|---|---|
| | Single-DC" Element (1,1) | Single-FAC" Element (1,2) | Single-MAC" Element (3,6) | Single-LAC" Element (8,8) |
| Thr/472*499, 21.9KB | 13.94755574 | 30.95834364 | 35.43984121 | 35.43984121 |
| AA/459*442, 21.9KB | 17.03133888 | 33.03741699 | 34.46826703 | 34.46826703 |
| AB/512*512, 22KB | 15.8030363 | 33.01288188 | 34.47099677 | 34.47099677 |
| AC/512*512, 340KB | 16.07762777 | 32.27712962 | 33.79547814 | 33.79547814 |
| Flu/348*288, 26KB | 13.22697662 | 32.74152234 | 34.53826509 | 34.53826509 |
| Gastro/640*480, 47.1KB | 13.69968559 | 31.38404996 | 35.46141742 | 35.46141742 |

Table 4. SSIM for the proposed approach

| Images/Size/ JPEG | SSIM of the Proposed Approach | | | |
|---|---|---|---|---|
| | Single-DC" Element (1,1) | Single-FAC" Element (1,2) | Single-MAC" Element (3,6) | Single-LAC" Element (8,8) |
| Thr/472*499, 21.9KB | 0.082951481 | 0.895439976 | 0.91813494 | 0.77977623 |
| AA/459*442, 21.9KB | 0.540008481 | 0.981986299 | 0.962351614 | 0.852759314 |
| AB/512*512, 22KB | 0.582614558 | 0.965833202 | 0.978976003 | 0.939601317 |
| AC/512*512, 340KB | 0.732765203 | 0.988080269 | 0.992311745 | 0.977517881 |
| Flu/348*288, 26KB | 0.4468566 | 0.976415496 | 0.988063836 | 0.989174577 |
| Gastro/640*480, 47.1KB | 0.51259705 | 0.976415496 | 0.983385471 | 0.867343767 |

Table 5. UIQI for the proposed approach

| Images/Size/ JPEG | UIQI for the Proposed Approach | | | |
|---|---|---|---|---|
| | Single-DC" Element (1,1) | Single-FAC" Element (1,2) | Single-MAC" Element (3,6) | Single-LAC" Element (8,8) |
| Thr/472*499, 21.9KB | 0.491369955 | 0.991868044 | 0.997112971 | 0.994662007 |
| AA/459*442, 21.9KB | 0.597094119 | 0.981986299 | 0.98718123 | 0.981198615 |
| AB/512*512, 22KB | 0.314558158 | 0.990700629 | 0.993380475 | 0.99140284 |
| AC/512*512, 340KB | 0.364645802 | 0.974713458 | 0.982441549 | 0.98155607 |
| Flu/348*288, 26KB | 0.153009019 | 0.989964331 | 0.993385972 | 0.99331525 |
| Gastro/640*480, 47.1KB | 0.343672644 | 0.992011314 | 0.996884846 | 0.994923823 |

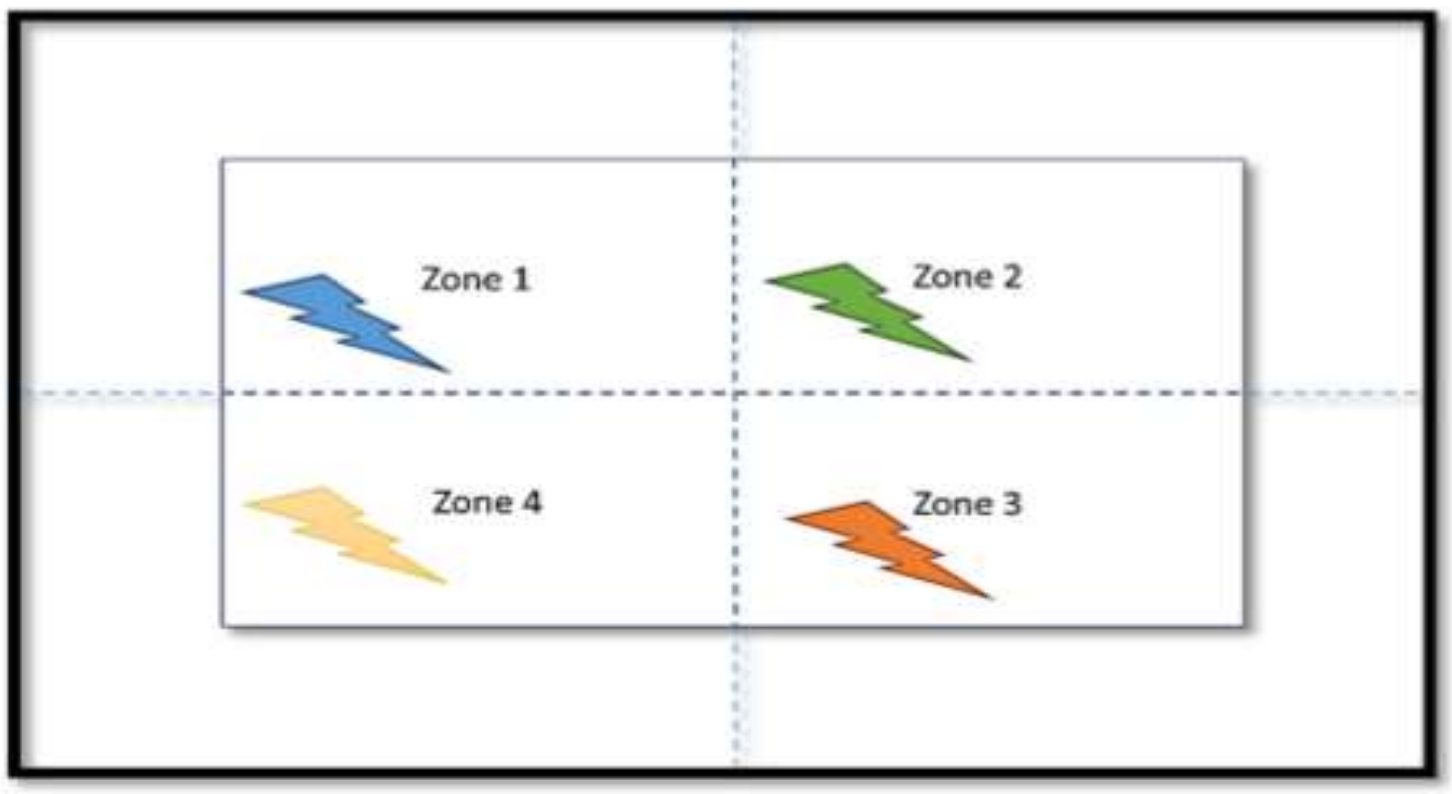

Figure 8. Infected component and locations of active attack for received RGB images.

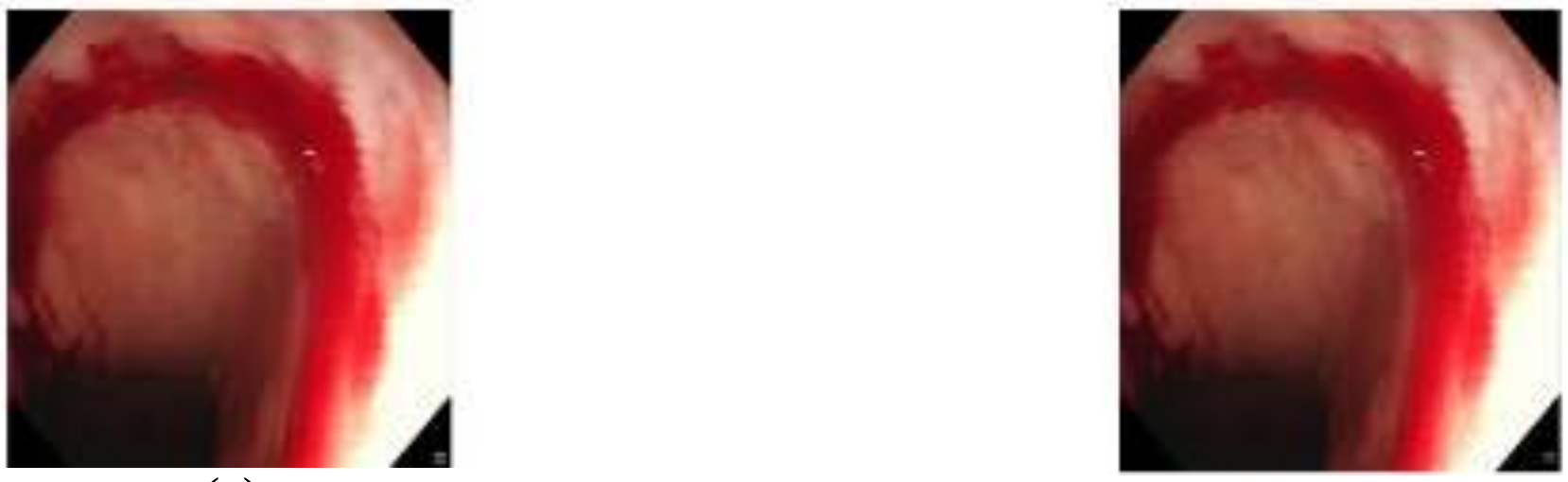

(a) (b)

Figure 9. (a) AB modified at red the by same blue component from raw 238 - 241 and column 300 – 303, (b) modified at red the by same blue component from raw 238 - 241 and column 300 - 303

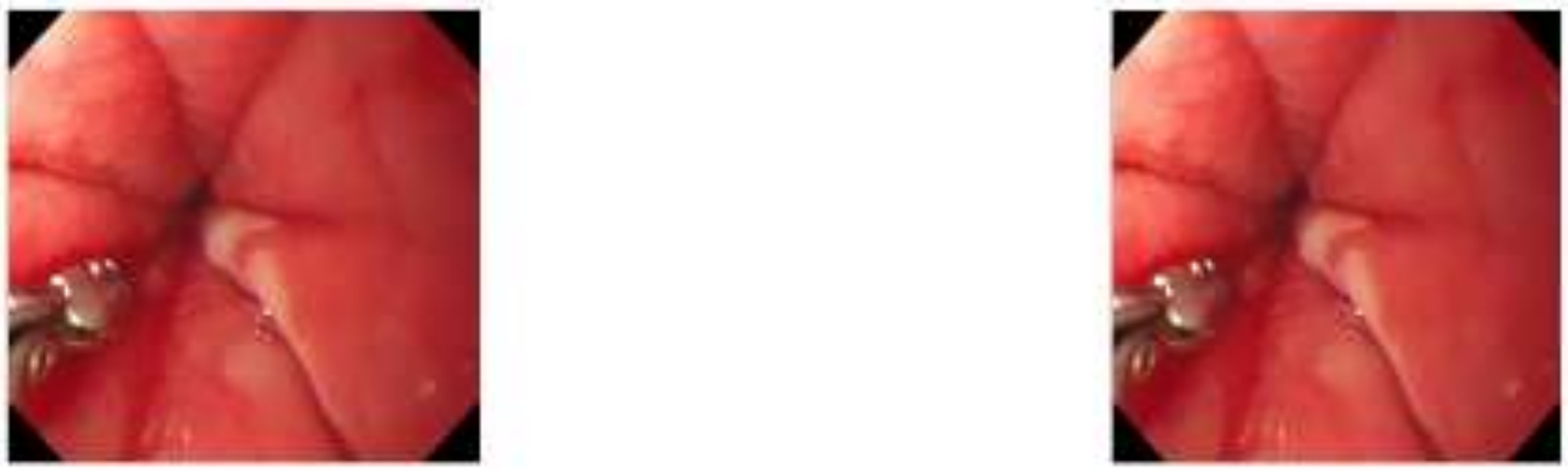

(a) (b)

Figure 10. (a) AC modified at blue by 255 from raw 238 - 241 and column 300 – 303, (b) AC modified at blue by the same of green component from raw 238 - 241 and column 300 – 303

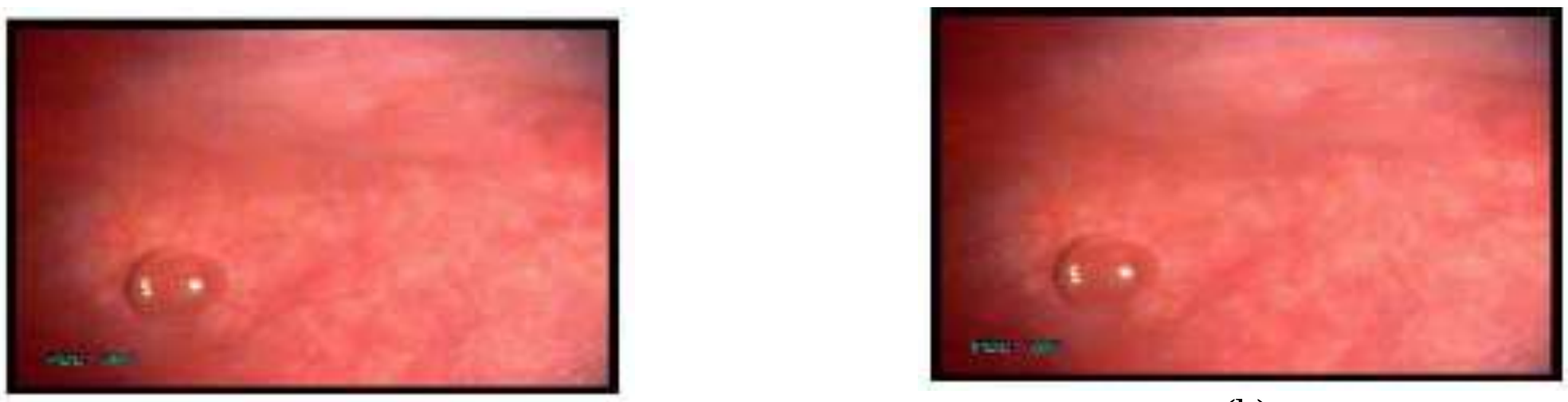

(a) (b)

Figure 11. (a) Flu modified at blue by green component at row from 138 - 141 and column 200 – 203, (b) Flu modified at red by the same blue component at row from 238 - 241 and column 300 - 303

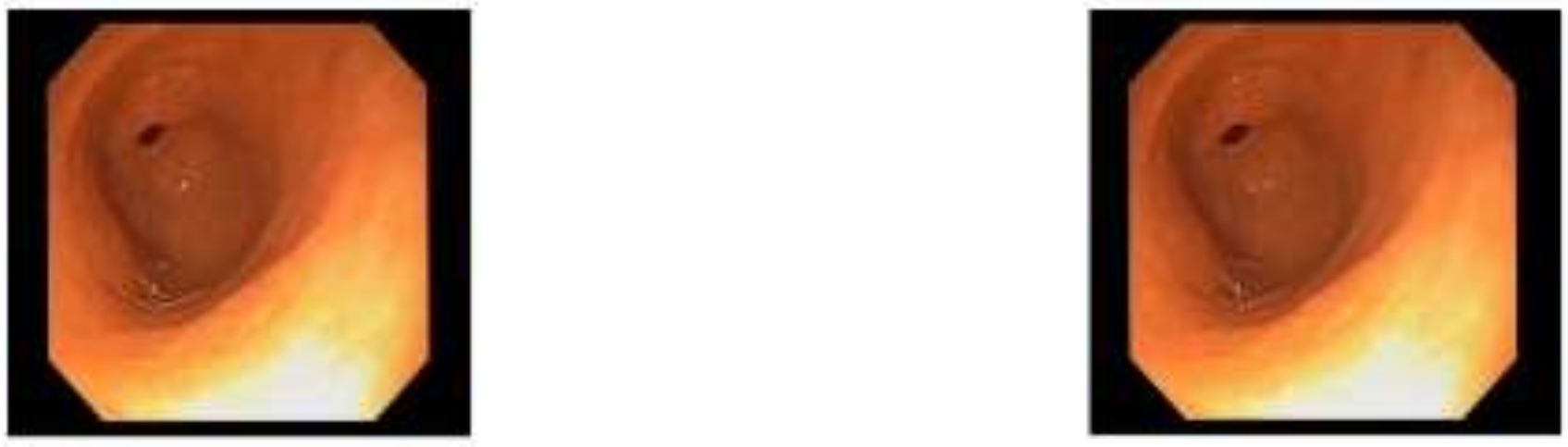

(a) (b)

Figure 12. (a) Gastro modified at red component, (b) Gastro modified at red component by 255 in raw 238 to 241 and column 300 to 303

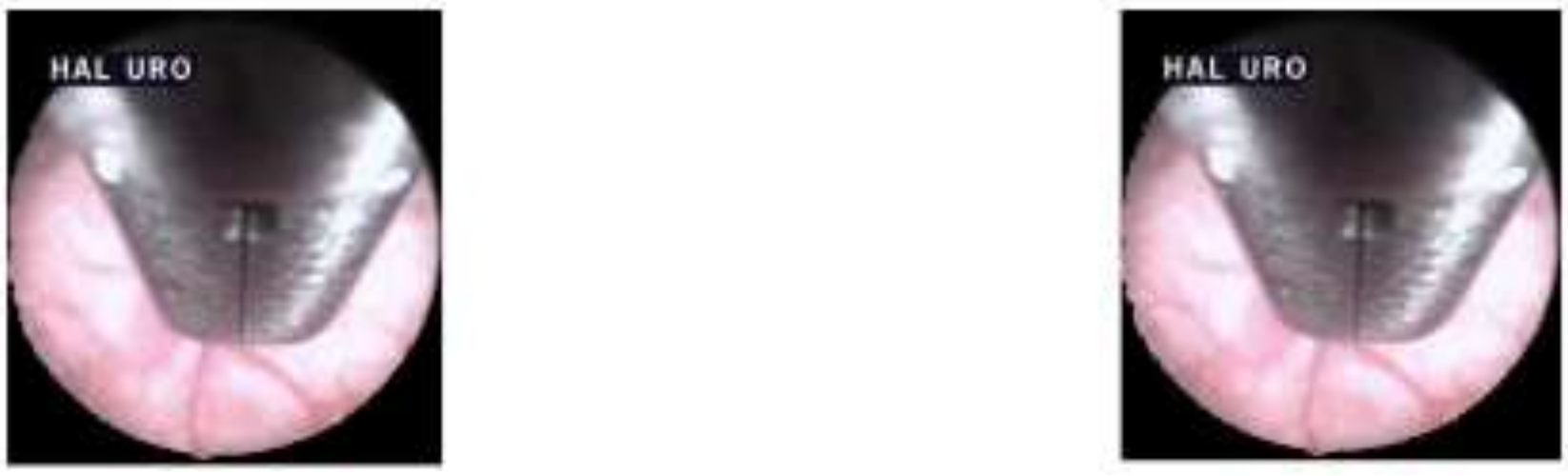

(a) (b)

Figure 13. (a) Thr modified at blue by green component at row from 138 - 141 and column 200 – 203, (b) Thr modified at blue component by 255 at row from 238 - 241 and column 200 - 203

## 4. CONCLUSION

This paper has proposed a new watermarking algorithm for medical images called Originality Identifier Code (OIC). The advantages of this code are that it has dynamic properties not fixed one for all images. It is also embedded in the digital images with the manner that is not noticeable while it is distributed on every pixel of the whole digital image. The power of the proposed algorithm arises from the fact that the encryption is distributed overall the entire image not only over the region of noninterest (RONI). At the same time, it preserves the features of the original images, which offers high protection for the patient privacy. The performance of the proposed algorithm has been evaluated using different performance measures. The Mean Square Error (MSE) of single MAC element (3, 6) gave the smallest value over all the other elements. The Mean Absolute Error (MAE) of single MAC element (3, 6) gave the smallest value over all the other elements. The Peak Signal to Noise Ratio (PSNR) of single MAC element (3, 6) gave the highest value over all the other elements. The Universal Image Quality Index (UIQI) of single MAC element (3, 6) gave the highest correlation value over all the other elements. the Structural Similarity Index (SSIM) of single MAC element (3, 6) shows that the high similarity between the original image and the encrypted one. From these reading it is clear that the choosing of single MAC element (3, 6) gave the best performance over the rest of elements. The proposed algorithm has a robust performance in detecting the intentional attacks that change the patient's diagnosis. And at the same time a high efficiency to restore the original image unlike most of the existing algorithms.

## REFERENCES

[1] Annaby, M. H., Rushdi, M. A., and Nehary, E. A., “Color image encryption using random transforms, phase retrieval, chaotic maps, and diffusion,” *Optics and Lasers in Engineering*, Vol. 103, pp. 9-23, Apr 2018.

[2] Abdel-Nabi, H., Al-Haj, A., “Medical imaging security using partial encryption and histogram shifting watermarking,” *Information Technology (ICIT), 2017 8th International Conference on,* IEEE., pp. 802–807, 2017.

[3] Shankar, K., and Eswaran, P., “RGB based multiple share creation in visual cryptography with aid of elliptic curve cryptography,” *China Commun.* Vol. 14, No. 2, pp. 118–130, 2017.

[4] Shankar, K., and Eswaran, P., “An efficient image encryption technique based on optimized key generation in ECC using genetic algorithm,” *Artificial Intelligence and Evolutionary Computations in Engineering Systems*. Springer, New Delhi. pp. 705–714, 2016.

[5] Kuang LQ, Zhang Y, Han X, “A Medical image authentication system based on reversible digital watermarking,” *Information Science and Engineering (ICISE), 2009 1st International Conference*. pp 1047–1050, 2009.

[6] Bhatnagar G, Jonathan WU QM, “Biometrics inspired watermarking based on a fractional dual tree complex wavelet transform,” *FuturGener Comput Syst*, Vol. 29, No. 1, pp. 182–195, 2013.

[7] Preeti Aggarwal, Renu Vig, Sonali Bhadoria, and C.G.Dethe , (September) “Role of Segmentation in Medical Imaging: A Comparative Study”, *International Journal of Computer Applications*, Volume 29– No.1, 2011.

[8] Zhen Ma, Joao Manuel, R. S. Tavares, R. M. Natal Jorge, “A review on the current segmentation algorithms for medical images”, *1st International Conference on Imaging Theory and Applications (IMAGAPP)*, Lisboa, Portugal, INSTICC Press, pp. 135-140, 2009.

[9] Liu, Jing, et al. "A Robust Multi-Watermarking Algorithm for Medical Images Based on DTCWT-DCT and Henon Map." *Applied Sciences*, Vol. 9, Vol. 4, 700, 2019.

[10] Nabil, E., “A modified flower pollination algorithm for global optimization,” *Expert Syst. Appl*. Vol. 57, pp. 192–203, 2016.

[11] Avudaiappan, T., et al. "Medical image security using dual encryption with oppositional based optimization algorithm." *Journal of medical systems*, Vol. 42, No. 11, 208, 2018.

[12] Roček, Aleš, et al. "Reversible watermarking in medical imaging with zero distortion in ROI,” *2017 24th IEEE International Conference on Electronics, Circuits and Systems (ICECS).* IEEE, 2017.

[13] Sharma, Abhilasha, Amit Kumar Singh, and S. P. Ghrera. "Secure hybrid robust watermarking technique for medical images," *Procedia Computer Science,* Vol. 70, pp. 778-784, 2015.

[14] Atta-ur-Rahman; Mahmud, M.; Sultan, K.; Aldhafferi, N.; Alqahtani, A.; Musleh, D. "Medical Image Watermarking for Fragility and Robustness: A chaos, error correcting codes and redundant residue number system based approach," *J. Med. Imaging Health Inform.*Vol. 8, pp. 1192–1200, 2018.

[15] Abokhdair, N.O.; Abd Manaf, A.; Alfagi, A.; Ab Sultan, M.; Mousavi, S.M.; Abd Manaf, Z.; Mohamad, F.S. "Patient data hiding and integrity control using prediction-based watermarking for brain MRI and CT scan images," *J. Med. Imaging Health Inform.*Vol 8, 691–702, 2018.

[16] Parah, S.A.; Sheikh, J.A.; Ahad, F.; Loan, N.A.; Bhat, G.M. "Information hiding in medical images: A robust medical image watermarking system for e-healthcare," *Multimed. Tools Appl.* Vol. 76, pp. 10599–10633, 2017.

[17] Thanki, R.; Borra, S.; Dwivedi, V.; Borisagar, K. "A roni based visible watermarking approach for medical image authentication," *J. Med. Syst.* Vol. 41, 143, 2017.

[18] Zhang, Z.;Wu, L.; Li, H.; Lai, H.; Zheng, C. "Dual watermarking algorithm for medical image," *J. Med. Imaging Health Inform*., Vol 7, pp. 607–622, 2017.

[19] Taher, F.; Kunhu, A.; Alahmad, H. "A new hybrid watermarking algorithm for MRI medical images using DWT and hash functions," *Proceedings of the Annual International Conference of the IEEE Engineering in Medicine and Biology Society*, Orlando, FL, USA, 16–20, pp. 1212–1215, August 2016.

[20] Zear, A.; Singh, A.K.; Kumar, P. "A proposed secure multiple watermarking technique based on DWT, DCT and SVD for application in medicine," *Multimed. Tools Appl*. 2018, 77, 4863–4882.

[21] Thanki, R.; Borra, S.; Dwivedi, V.; Borisagar, K. "An efficient medical image watermarking scheme based on FDCuT-DCT," *Eng. Sci. Technol*. 2017, 20, 1366–1379.

[22] Wang, J.W.; Lian, S.G.; Shi, Y.Q. "Hybrid multiplicative multi-watermarking in DWT domain," *Multidimens. Syst. Signal Process.*, Vol 28, pp. 617–636, 2017.

[23] Yuan, X.C.; Li, M. "Local multi-watermarking method based on robust and adaptive feature extraction," *Signal Process.*, Vol. 149, pp. 103–117, 2018.

[24] Singh, A.K.; Kumar, B.; Dave, M.; Mohan, "A multiple watermarking on medical images using selective discrete wavelet transform coefficients," *J. Med. Imaging Health Inform.*Vol. 5, 607–614, 2015.

[25] Arsalan, M.; Qureshi, A.; Khan, A.U.; Rajarajan, M. "Protection of medical images and patient related information in healthcare: Using an intelligent and reversible watermarking technique," *Appl. Soft Comput.* Vol 51, pp. 168–179, 2017.

[26] Vishakha, K.; Kushal, T.; Hitesh, N. "Novel variants of a histogram shift-based reversible watermarking technique for medical images to improve hiding capacity," *J. Healthc. Eng.* Vol. 6, 2017.

[27] Turuk, M.P.; Dhande, A.P. "A novel reversible multiple medical image watermarking for health information system," *J. Med. Syst.* Vol. 40, pp. 269–279, 2016.

[28] Lei, B.; Tan, E.L.; Chen, S.; Ni, D.; Wang, T.; Lei, H. "Reversible watermarking scheme for medical image based on differential evolution," *Expert Syst. Appl.* Vol. 41, pp. 3178–3188, 2014.

[29] Abokhdair, N.O.; Abd Manaf, A.; Alfagi, A.; Ab Sultan, M.; Mousavi, S.M.; Abd Manaf, Z.; Mohamad, F.S. "Patient data hiding and integrity control using prediction-based watermarking for brain MRI and CT scan images," *J. Med. Imaging Health Inform.* Vol. 8, pp. 691–702, 2018.

[30] Parah, S.A.; Sheikh, J.A.; Ahad, F.; Loan, N.A.; Bhat, G.M. "Information hiding in medical images: A robust medical image watermarking system for e-healthcare," *Multimed. Tools Appl.*, Vol. 76, 10599–10633, 2017.

[31] Thanki, R.; Borra, S.; Dwivedi, V.; Borisagar, K. "A roni based visible watermarking approach for medical image authentication," *J. Med. Syst.*, Vol. 41, 143, 2017.